\documentclass[twocolumn,showpacs,tightenlines,nofootinbib,prc]{revtex4}
\usepackage[utf8]{inputenc}
\usepackage{amsmath,amssymb,graphicx}
\DeclareMathOperator\arctanh{arctanh}
\usepackage{placeins}
\usepackage[none]{hyphenat}
\usepackage[usenames,dvipsnames]{color}
\usepackage{hyperref}
\usepackage{ulem}
\hypersetup{colorlinks=true,citecolor=Blue,linkcolor=Blue,urlcolor=Blue}
\begin{document}

\title{\large{Supersymmetric inversion of effective-range expansions}}
\author{Bikashkali Midya}
\email{bikash.midya@gmail.com}
\author{Jérémie Evrard}
\author{Sylvain Abramowicz}
\author{O. L. Ramírez Suárez}
\author{Jean-Marc Sparenberg}
\email{jmspar@ulb.ac.be}
\affiliation{Physique nucléaire et Physique quantique, École polytechnique de Bruxelles, Université libre de Bruxelles (ULB), CP 229,
B-1050 Brussels, Belgium. \vspace{.2 cm}}

\begin{abstract}
A complete and consistent inversion technique is proposed to derive an accurate interaction potential from an effective-range function for a given partial wave in the neutral case. First, the effective-range function is Taylor or Padé expanded, which allows high precision fitting of the  experimental scattering phase shifts with a minimal number of parameters on a large energy range. Second, the corresponding poles of the scattering matrix are extracted in the complex wave-number plane. Third, the interaction potential is constructed with supersymmetric transformations of the radial Schrödinger equation. As an illustration, the method is applied to the experimental phase shifts of the neutron-proton elastic scattering in the $^1S_0$ and $^1D_2$ channels on the $[0-350]$ MeV laboratory energy interval. 

\end{abstract}

\pacs{ 03.65.Nk, 13.75.Cs, 21.45.-v, 11.30.Pb\vspace{.1 cm}}
\maketitle 
\section{Introduction}

In quantum mechanics, the simplest scattering experiment is the elastic collision between two spinless particles \cite{taylor:72}.
The measured quantity is the elastic-scattering differential cross section,
which is the squared modulus of the scattering amplitude.
In principle,
the scattering amplitude depends continuously on two parameters:
the deflection angle $\theta$ and the energy $E$.
Thanks to rotational invariance,
the continuous dependence on $\theta$ can however be replaced by a discrete sum on partial waves,
with a number of significant terms and hence a complexity of the $\theta$ dependence increasing with energy.
This well-known expansion strongly simplifies both the theoretical description and the experimental measure of the collision.
Typically, 9 partial waves are sufficient to describe the elastic scattering of the nucleon-nucleon system below the first inelastic threshold (pion production), i.e. on the $[0-350]$ MeV energy interval in the laboratory frame \cite{stoks:93}.

Less known is the fact that the continuous energy dependence can also be replaced by a discrete sum,
leading to a similar simplification.
This is made possible by the use of the effective-range function,
a function directly related to the partial-wave scattering matrix or phase shift,
which can be series expanded as a function of energy.
The usual effective-range expansion [see Eq.\ (\ref{ERE}) below] is a Taylor expansion,
which is only valid at low energy and hence of reduced interest to parametrize scattering cross sections in all generality.
Typically, for the nucleon-nucleon $S$ wave,
this expansion is usually believed to be useful to fit experimental phase shifts up to 5 MeV only \cite{deswart:95}.
It was however realized by several authors \cite{hartt:81,kievsky:97,babenko:05,pupasov:11}
that a Padé approximant or rational expansion of the effective-range function
was much more appropriate than a Taylor expansion to parametrize data efficiently.
For instance, as shown below, for the nucleon-nucleon $S$ wave,
a Padé expansion of order [3/2], which depends on 6 parameters, fits the effective-range function on the whole elastic-scattering region.
This number of parameters is decreasing with increasing partial wave,
as a consequence of the centrifugal-barrier effect.

Combining the partial-wave decomposition with Padé approximants of the effective-range functions
is thus an extremely efficient way of parameterizing experimental scattering data with high precision.
It can actually be seen as an optimal multi-energy phase-shift analysis.
Another striking advantage of this method, at least in the neutral case,
is that it also leads to an exactly-solvable potential model for each partial wave to which it is applied.
It can thus be considered as an inverse-scattering technique \cite{chadan:89}
since it allows to generate an interaction potential from the measured scattering data.
Indeed, a rational expansion of the effective-range function
leads to a rational expansion of the partial-wave scattering matrix
as a function of the wave number [see Eq.\ (\ref{SK}) below].
Such an expansion leads in turn to scattering-matrix poles,
which are the basic ingredient for the inversion technique based on supersymmetric quantum mechanics
\cite{sparenberg:97a,baye:04,baye:14}. 
This technique is related to the better-known inversion methods attributed to integral equations with separable kernels \cite{chadan:89}, which were
also applied to the nucleon-nucleon system \cite{Kuberczyk:87,Kirst:89}, but with much larger numbers of poles. The present method admits the additional advantage of leading to exactly-solvable potentials with compact analytical expressions. These potentials can also be seen as generalizations of the Bargmann potentials \cite{Bargmann:49} as obtained through Darboux transformations for the nucleon-nucleon system \cite{Leeb:92}.

The method presented here is thus complete, in a sense, as it allows to start from experimental cross sections,
to parametrize them in an extremely efficient way and to build the corresponding interaction potentials in an elegant mathematical form. In the following, we first elaborate the key ingredients of the scattering theory and Padé/Taylor expansion method of effective-range function, then describe the radial supersymmetric inversion technique, and finally apply our method to the neutron-proton system in the singlet states, which is neutral and for which spins can be neglected.

\section{Supersymmetric inversion of effective-range function}

Let us consider a collision between two spin-0 particles
with a center-of-mass energy $E=\hbar^2 k^2/2\mu$,
where $\hbar$ is the reduced Planck constant, $k$ is the wave number and $\mu$ is the reduced mass of the particles (in the following, we choose the reduced units $\hbar = 2 \mu =1$).
Each partial wave is associated with an orbital angular-momentum quantum number $l$,
which is not written explicitly in the following as each partial wave is treated independently.
A partial wave is characterized by a scattering phase shift $\delta(k)$
or a unitary partial-wave scattering matrix $S(k) = e^{2i \delta(k)}$, 
which continuously depend on the energy or wave number.
The effective-range function is defined as
\begin{equation}
K(k^2) \equiv k^{2l+1} \cot \delta(k) = i k^{2l +1} \frac{S(k)+1}{S(k)-1},
\label{Kl}
\end{equation}
with a power of $k$ depending on the partial wave.
Under rather general assumptions, this function can be proven to be analytic at low energy \cite{newton:82}.
It thus admits the so-called effective-range (Taylor) expansion,
\begin{equation}
K(k^2\rightarrow 0) = - \frac{1}{a} + \frac{r}{2} k^2 - P r^3 k^4 + O(k^6),
\label{ERE}
\end{equation}
where the scattering length $a$, the effective range $r$ and the shape parameter $P$
can be seen as a discrete set of parameters characterizing the elastic phase shift at small values of the continuum energy.

The effective-range function is actually meromorphic, i.e.\ it admits poles;
it is thus advantageous to replace expansion (\ref{ERE}) by a Padé approximant of order $[M/N]$,
\begin{equation}
 K(k^2) = \frac{P^{[M]}(k^2)}{Q^{[N]}(k^2)},
 \label{Krat}
\end{equation}
which recovers equation (\ref{ERE}) in the $N=0$ particular case. Moreover, the high energy $( k\rightarrow \infty )$ behavior of the phase-shifts \cite{taylor:72} $\delta(k) \sim  k^{-1}$ (modulo $\pi$) is satisfied by this functional form when $M-N = l+1$. Equation (\ref{Krat}) leads to an expansion of the scattering matrix as a rational function of $k$.
Indeed, inverting Eq.\ (\ref{Kl}) leads to
\begin{equation}
S(k) = \frac{K(k^2)+i k^{2l+1}}{K(k^2)-i k^{2l+1}},
\label{SK}
\end{equation}
with a power of $k$ depending on the partial wave.
Equation (\ref{SK}) shows that when the effective-range function is a polynomial or a rational function (of the energy),
the scattering matrix automatically becomes a rational function of $k$,
\begin{equation}
 S(k) = \prod_{j=0}^{n-1} \frac{k+i \kappa_j}{k-i \kappa_j},
 \label{Spol}
\end{equation}
with poles at $k=i \kappa_j$ satisfying
\begin{equation}
 P^{[M]}(-\kappa_j^2) - (-1)^{l+1} \kappa_j^{2l+1} Q^{[N]}(-\kappa_j^2) = 0.
 \label{PQ}
\end{equation}
Equation (\ref{PQ}) shows that these poles depend on the coefficients of the effective-range expansion
and satisfy the following properties:
\begin{enumerate}
 \item their number $n = \max(2M, 2N+2l+1)$;
 \item they are either purely imaginary or symmetric with respect to the imaginary axis,
 which warrants the unitarity of the scattering matrix;
 \item when $l>0$, they satisfy the conditions
 \begin{equation}
  \sum_{j=0}^{n-1} \frac{1}{\kappa_j^\alpha} = 0, \quad \alpha = 1, 3\dots (2l-1),
  \label{polcond}
 \end{equation}
 as can be seen by comparing the denominator of Eq.\ (\ref{Spol}) with Eq.\ (\ref{PQ}).
\end{enumerate}

When parametrizing experimental phase shifts, two approaches are possible to determine the scattering-matrix poles.
The first approach consists in finding the minimal orders $M$ and $N$ leading to a satisfactory effective-range function.
The poles are then deduced from the above equations.
The advantage is that they automatically satisfy conditions (\ref{polcond}).
The drawback of this approach is that these poles can be either imaginary or complex,
while complex poles sometimes lead to oscillations in the potentials deduced from supersymmetric quantum mechanics.
To avoid such oscillations, it is thus necessary to constrain the poles to stay on the imaginary $k-$axis \cite{samsonov:03}.
This can be achieved by directly fitting the phase shifts as
\begin{equation}
 \delta(k) = -\sum_{j=0}^{n-1} \arctan \frac{k}{\kappa_j},
 \label{deltasum}
\end{equation}
which is equivalent to Eq.\ (\ref{Spol}).
The drawback is then that the poles have to be constrained by conditions (\ref{polcond}) for $l>0$
in order for the effective-range function to be well defined.
In the following, both approaches will be used.

Equations (\ref{Spol}) and (\ref{deltasum}) are in fact associated to a chain of $n$ supersymmetry transformations \cite{sparenberg:97a,baye:04,baye:14} of the radial Schr\"odinger equations $H_j \psi \equiv - \psi''(k,r) + V_j(r) \psi(k,r) = k^2 \psi(k,r)$, with $j = 0,1,2,...,n$ and with a purely centrifugal initial potential $V_0(r) = l(l+1) r^{-2}$. Each first-order supersymmetry transformation $L_j = -d/dr + v_j'/v_j$ is an algebraic transformation which transforms the Hamiltonian $H_j$ of the chain to a new Hamiltonian $H_{j+1}$ with $V_{j+1} = V_j -  2 (v_j'/v_j)'$. Here, the factorization solutions, $v_j \equiv v(\kappa_j,r)$, are the solutions of $H_j$ corresponding to the distinct factorization energies $\varepsilon_j$ which are related to the scattering matrix poles $i\kappa_j$ by $\varepsilon_j = - \kappa_j^2$. The Hamiltonians $H_j$ and $H_{j+1}$ share identical spectral characteristics, whereas each successive transformation of the chain modifies the phase-shift by subtracting an $\arctan(k/\kappa_j)$ term from the phase-shift of the former Hamiltonian.  The compact expression of the final potential of the chain can readily be expressed by the following Crum-Krein formula \cite{crum:1955,krein:1957}
\begin{equation}
V_{n} = \frac{l(l+1)}{r^2} - 2\frac{d^2}{dr^2} \ln W[u_0,u_1,...,u_{n-1}], \label{pot}
\end{equation}
where $u_j \equiv u(\kappa_j,r)$ are the solutions of the initial Schr\"odinger equation $-u_j'' + l(l+1) r^{-2} u_j = -\kappa_j^2 u_j$. When $u_j$ is associated to a pole which lies in the upper (lower) half $k-$plane and is regular at the origin and exponentially increasing at infinity (respectively singular at the origin and exponentially decreasing at infinity), it is characterized as the left- (respectively right-) regular solutions. Each supersymmetry transformation corresponding to these solutions is responsible for the increment (decrement) of the potential singularity at the origin by one unit. Since $V_{0}$ has a repulsive core of the form $l(l+1)/r^2$, the singularity strength of the final potential is therefore equal to $l$ plus the difference of the number of left-regular minus right-regular transformations. Thus the aforementioned two types of factorization solutions are the key ingredients to build a potential with singularity at the origin. 

For the $l=0$ partial wave, the solutions of the Hamiltonian $H_{0}$ are given by $u_j \propto \sinh \kappa_j r$, or $\exp(\kappa_j r)$. The first (second) solution corresponds to the left- (right-) regular transformation if $\operatorname{Re}(\kappa_j) > 0$ (respectively $<0$). On the other hand, for $l>0$, the solutions $u_j$ of the purely centrifugal potential can be obtained by applying $l$ zero-energy transformations on the above mentioned $l=0$ left and right regular solutions \cite{baye:14}.

\onecolumngrid
\begin{center}
\begin{figure}[h]
\includegraphics[width=8cm,height=5cm]{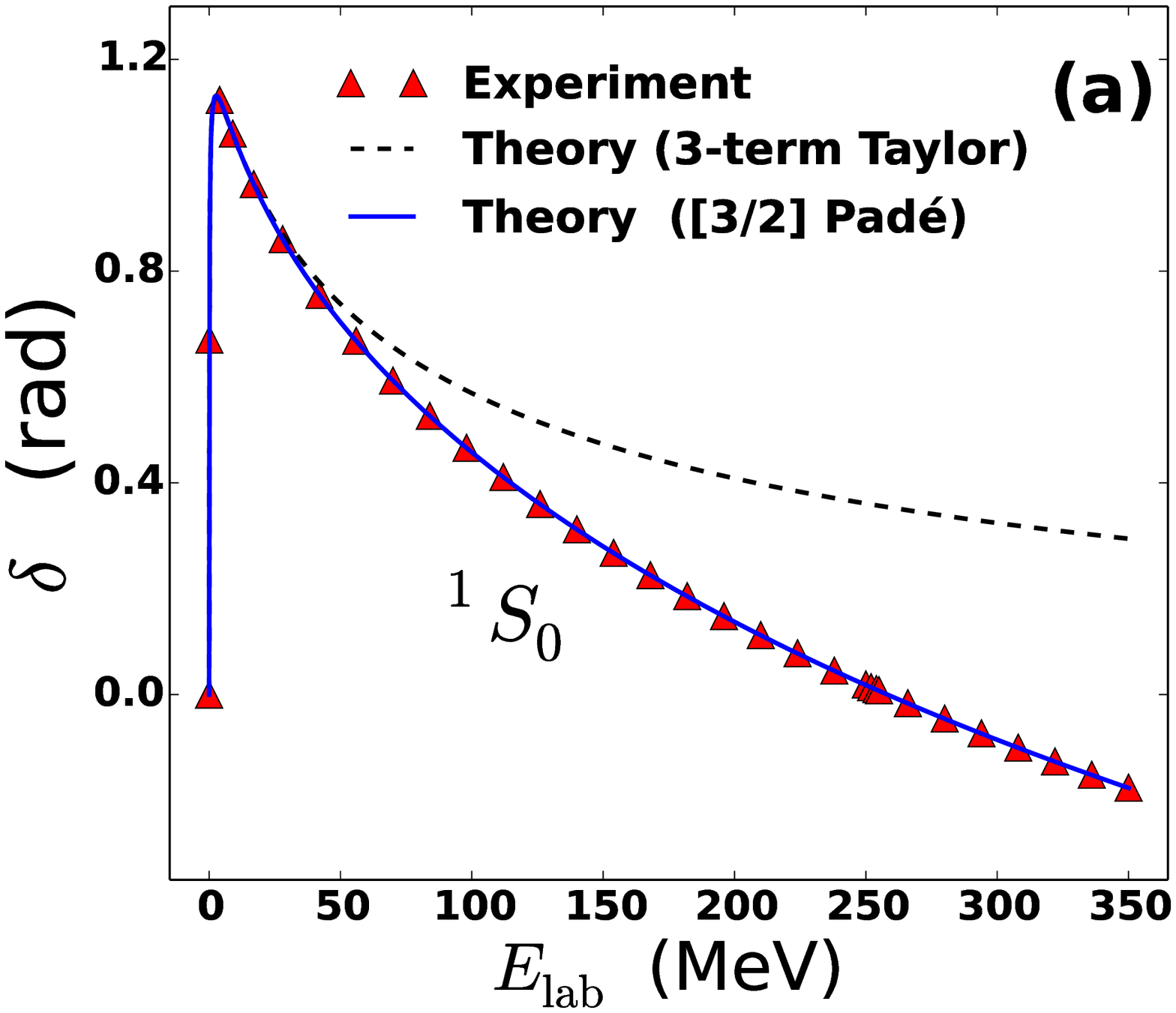}
\includegraphics[width=7cm,height=4.85cm]{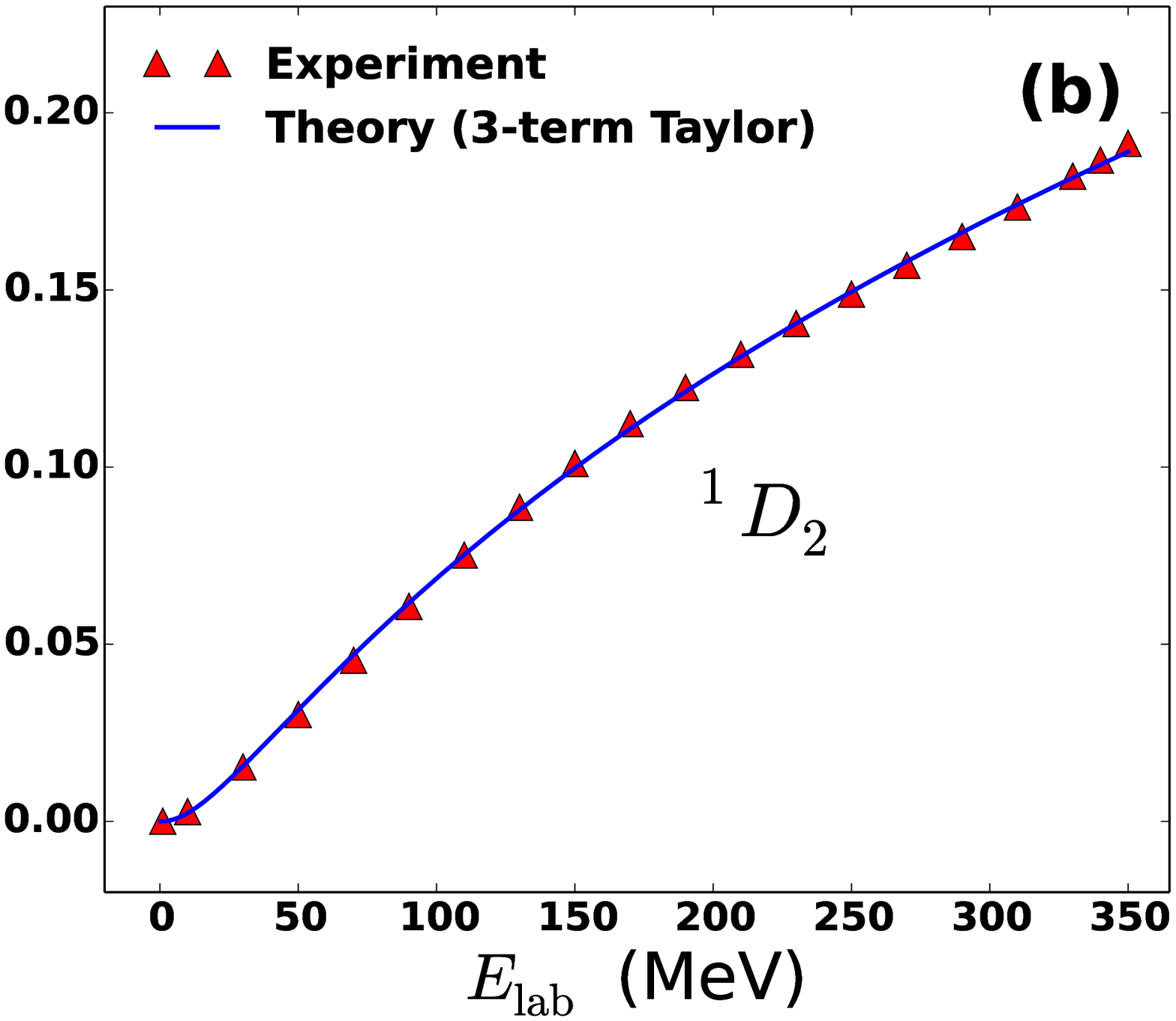}
 \caption{\label{fig:delta} (Color online) (a) Experimental neutron-proton singlet $S$-wave phase shifts \cite{stoks:93} and theoretical phase shifts [equation (\ref{deltasum})]  deduced from effective-range-function fits. (b) Same as in (a) but for singlet $D$-wave.}
\end{figure}
\begin{figure}[h]
\includegraphics[width=7.75cm,height=5cm]{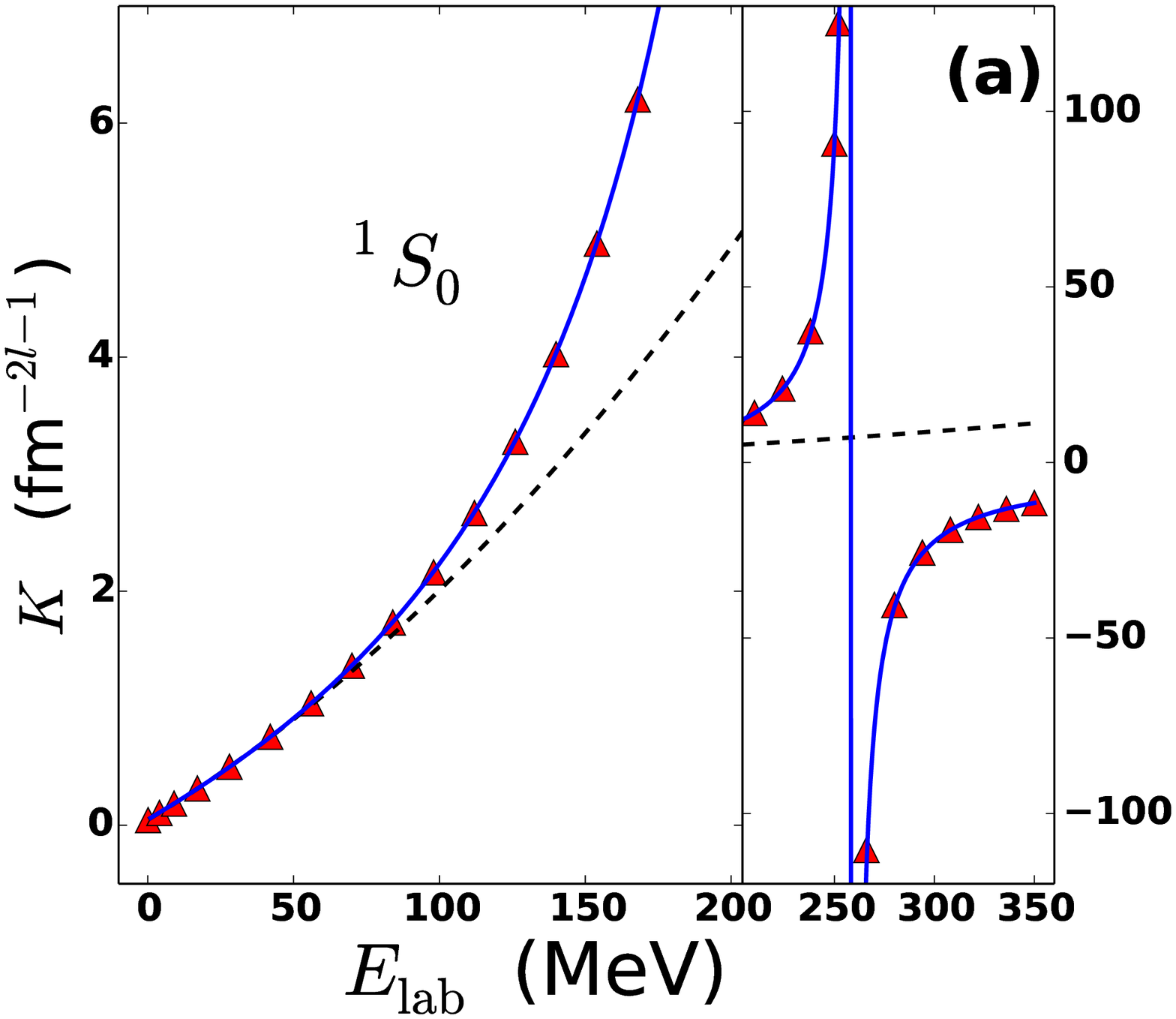} 
\includegraphics[width=7.5cm,height=4.9cm]{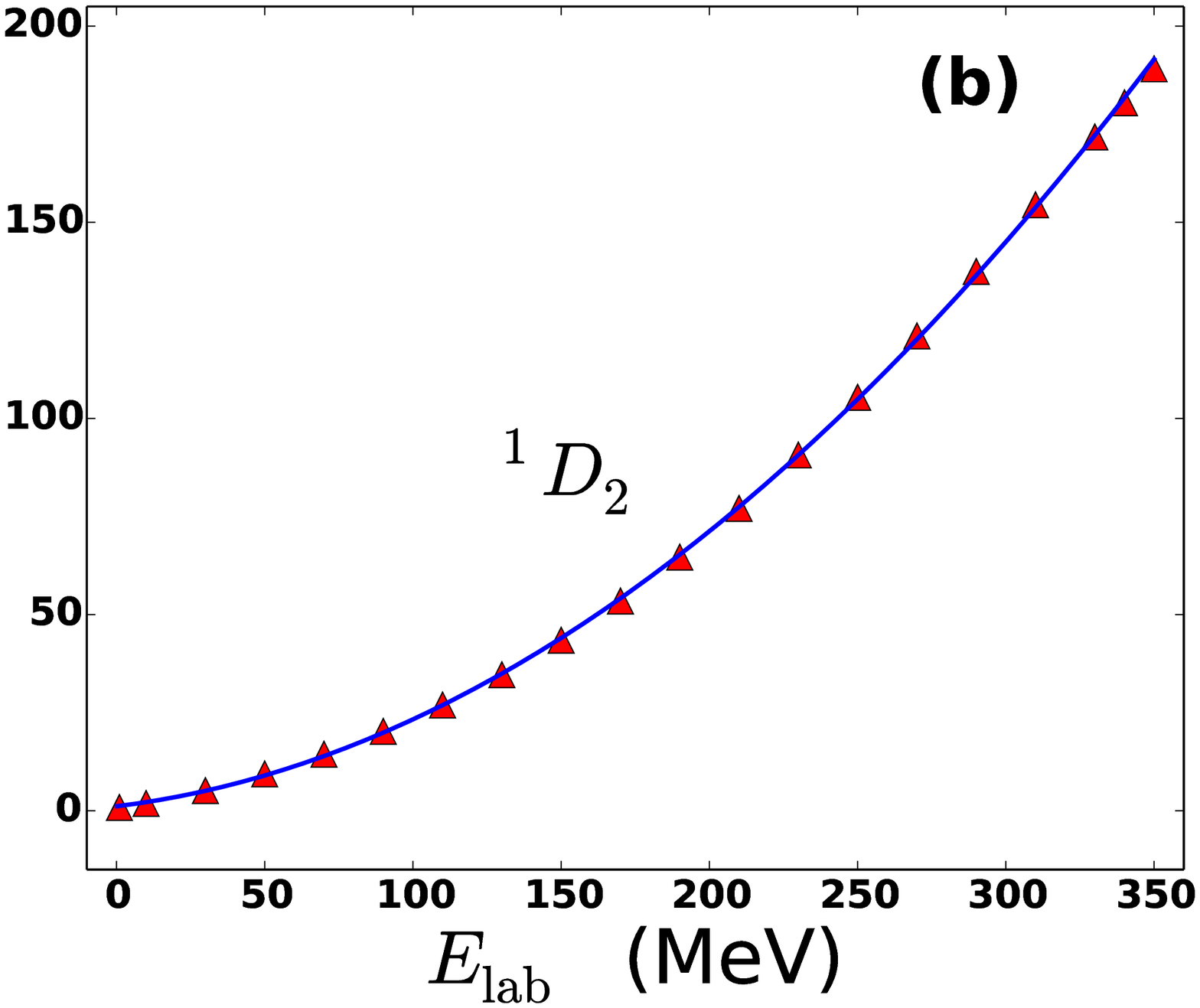}
 \caption{\label{fig:erf} (Color online)  Effective-range functions corresponding to the phase shifts of Fig.\ \ref{fig:delta}.
 (a) S-wave: experiment \cite{stoks:93}, 3-term Taylor expansion, [3/2] Padé expansion [left (resp. right) axis is used for the left (resp. right) part of the figure]; (b) D-wave: experiment \cite{stoks:93}, 3-term Taylor expansion.}
\end{figure}
\end{center}
\twocolumngrid

\section{Application to neutron-proton elastic scattering}

Let us now apply our method to the neutron-proton singlet-state phase shifts deduced from elastic scattering experimental data \cite{stoks:93}
on the $[0-350]$ MeV laboratory energy interval.
Note here that the laboratory energy and the center-of-mass momentum squared $k^2$ in non-relativistic kinematics are related by $E_{\mathrm{lab}} = \frac{\hbar^2}{2\mu}\frac{m_p+m_n}{m_p} k^2,$ where $m_n= 939.565$ MeV and $m_p = 938.272$ MeV are the mass of a neutron and proton, respectively and $\hbar c = 197.33$ MeV fm (a relativistic treatment \cite{stoks:93} has a negligible impact on the results).

\begin{figure}[htb]
  \centering
 \includegraphics[width=7.8cm,height=4.5cm]{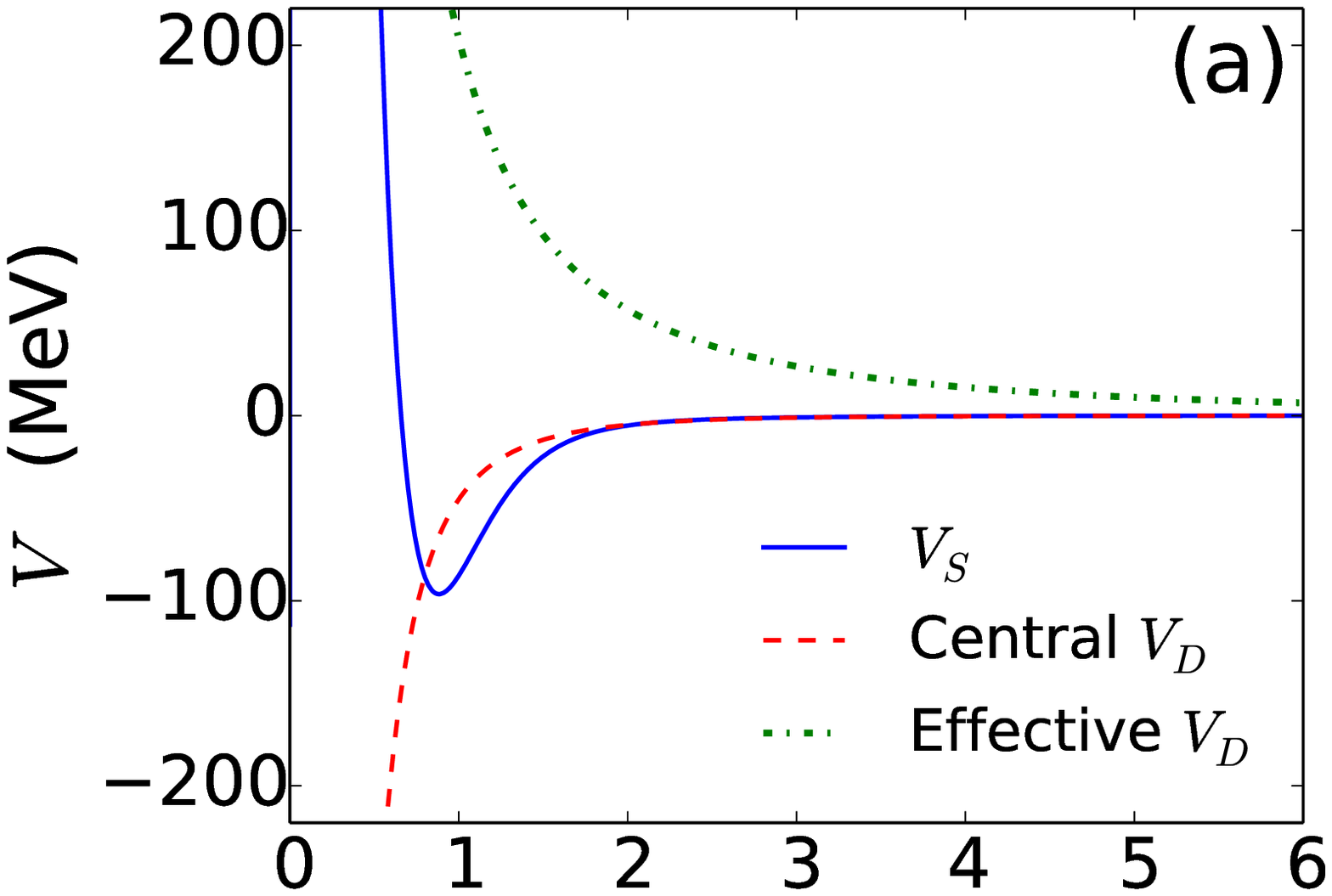}
 \vspace{-.35cm}
 
  \includegraphics[width=8cm,height=5cm]{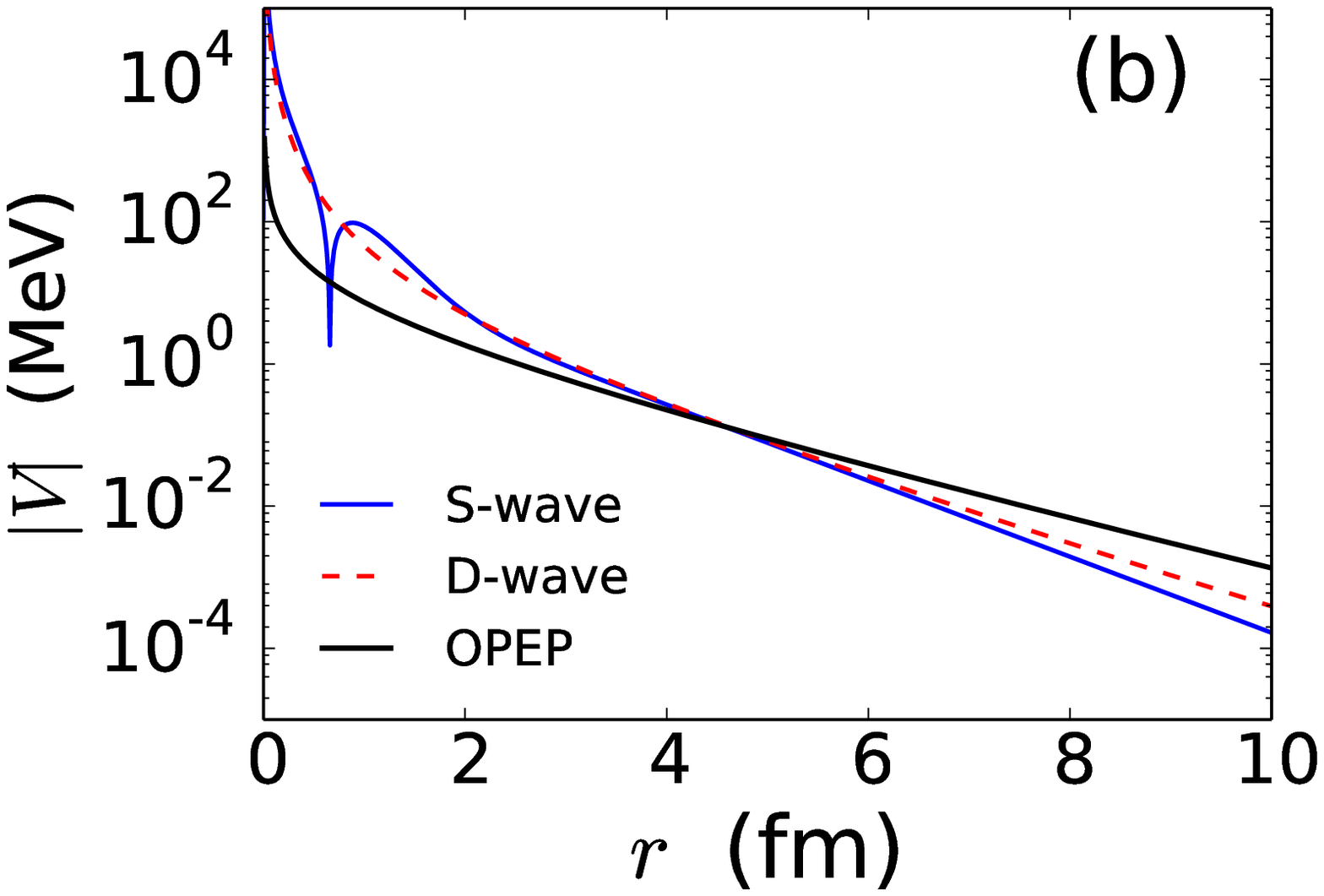}
 
 \caption{\label{fig:pot} (Color online) (a) Neutron-proton inversion potentials for the singlet $S$- and $D$-waves (central and effective potentials). (b) Plot of the asymptotic behavior of the central S- and D-wave potential together with one pion exchange potential (OPEP) in logarithmic scale. Both figures are plotted after multiplying $\hbar^2/(2\mu) = 41.47$ MeV fm$^2$ to the corresponding potentials. }
\end{figure}

First, let us revisit the $l=0$ results, for which no conditions on the poles apply. Hence, the simplest method is based on the direct fit of the phase shifts with Eq.\ (\ref{deltasum}).
In Ref.\ \cite{sparenberg:97a},
a five-pole fit of the data was found satisfactory but two poles were complex,
which led to a small oscillation in the potential.
This default was fixed in Ref.\ \cite{samsonov:02},
where a satisfactory fit with six poles was found by constraining them to lie on imaginary axis. The quality of this fit, by the poles $i \kappa_j$ with $\kappa_{0, . . . , 5} = -0.0401, -0.7540, 0.6152, 2.0424, 4.1650, 4.6$  fm$^{-1}$, can be seen on Fig. \ref{fig:delta}(a). The
corresponding effective-range function is associated with the following [3/2] Padé expansion 
\begin{equation}
K_S(k^2) = \frac{0.0422 + 1.3793 ~k^2 + 2.0105 ~k^4 -0.058 ~k^6}{1+ 1.5986~ k^2 -0.6164 ~k^4}
\end{equation}
which is shown in Fig \ref{fig:erf}(a). On the other hand, a 3-term Taylor expansion of the effective-range function $K_S(k^2) = 0.04219 + 1.30386 k^2 + 0.06883 k^4$ [also shown in Fig \ref{fig:erf}(a)], with scattering length $a= -23.7$ fm, effective range $r= 2.608$ fm, and with scattering matrix poles $\kappa_{0,..., 3} = -0.0401, -4.6917, 0.8365, 3.8953$ fm$^{-1}$, is able to fit the phase shifts up to 30 MeV lab energy only [as displayed in Fig \ref{fig:delta}(a)]. This shows the interest of using a Padé expansion rather than a Taylor expansion. Moreover, the order of the Padé expansion appropriately resembles the correct high energy behavior of the phase shifts (which is $-\pi$, as can be checked immediately). The corresponding interaction potential can be written in two equivalent forms
\begin{eqnarray}
 V_{S} & = & -2\frac{d^2}{dr^2}
 \ln W \big[e^{-\kappa_0 r}, e^{-\kappa_1 r}, \sinh(\kappa_2 r), \nonumber \\
 && \sinh(\kappa_3 r), \sinh(\kappa_4 r), \sinh(\kappa_5 r)\big] \\
 & = & -2\frac{d^2}{dr^2}
 \ln W \big[\cosh(\kappa_2 r + \alpha_{02} +\alpha_{21}), \nonumber \\
 && \sinh(\kappa_3 r +\alpha_{03}+\alpha_{13}), \sinh(\kappa_4 r + \alpha_{04} + \alpha_{14}), \nonumber \\
 &&  \sinh(\kappa_5 r + \alpha_{05} + \alpha_{15})\big],
\end{eqnarray}
with $\alpha_{ij} = \arctanh(\kappa_i/\kappa_j)$.
The potential is represented on Fig.\ \ref{fig:pot}; it displays both a correct one-pion-exchange asymptotic behaviour and a repulsive core at the origin, like standard nucleon-nucleon potentials.

Next, we consider the inversion of neutron-proton elastic scattering experimental phase shifts in the $^1D_2$ channel. For these data, the 3-term Taylor effective-range expansion (\ref{ERE}) is sufficient to fit the data with high precision,
as shown on Fig.\ \ref{fig:erf}(b).
The corresponding parameters read : $a =  0.88762$ fm, $r = 15.33061$ fm,  $P = -0.00246$ and the corresponding poles of the scattering matrix are
$\kappa_{0,...,4} = -0.4294, -0.8827, -8.7653, 0.7750,~0.4376 ~\mathrm{fm}^{-1}$. Remarkably, all these poles lie on the imaginary axis of the complex wave-number plane,
whereas this constraint was not imposed to them.
The sum of five arctangents corresponding to these five poles is plotted in Fig.\ \ref{fig:delta}(b),
which confirms the excellent quality of the fit with the experimental data. However, since the condition $M-N =l+1$ is not satisfied, the high energy behavior of the phase shift tends to $\pi/2$. 

The compact expressions of the corresponding effective potential $V_D(r)$ is given by Eq. (\ref{pot}), 
where right regular transformation functions $u_{0,1,2}$ and left regular solutions $u_{3,4}$ are associated with negative poles and positive poles, respectively and read
\begin{eqnarray*}
u_j(r) & = & \left(1+\tfrac{3}{\kappa_j r}+\tfrac{3}{\kappa_j^2 r^2}\right) e^{-\kappa_j r},\ j=0, 1, 2, \\
u_j(r) & = & \tfrac{3}{\kappa_j r} \cosh(\kappa_j r) - \left(\tfrac{3}{\kappa_j^2 r^2} + 1\right) \sinh(\kappa_j r), \ j=3, 4.
\end{eqnarray*}
In Fig.\ \ref{fig:pot}, we have plotted this potential, together with the corresponding central potential after extracting the centrifugal term. Clearly the central potential is a deep potential with attractive singular core.
Contrary to the $S$-wave potential, this potential belongs to the family of deep potentials,
as proposed by the Moscow group \cite{kukulin:99}.
This is due to the fact that the $D$-wave phase shifts are positive.
This also supports the results of Ref.\ \cite{sparenberg:02a},
where a parity-independent deep potential was obtained from $S$- and $P$-wave phase-shift inversion,
with the inclusion of Pauli forbidden states.
A question raised at that time was the incompatibility of this potential with the $D$- and $F$-waves,
hence the interest of directly inverting phase shifts for these waves.
Let us stress that because of the centrifugal term the $D$-wave potential obtained here is only constrained by data above $0.7$ fm.
Figure \ref{fig:pot} shows that even above this radius the $S$-wave potential is deeper than the $D$-wave one.
Hence, adding a forbidden bound state to the $S$-wave potential will probably not allow to fit the $D$-wave simultaneously. A similar conclusion was drawn in Ref.\ \cite{sprung:75}, which can be accounted by allowing explicit non-localities in the potential, for instance through a momentum dependence \cite{Korinek:96} or a quadratic dependence on angular momentum $\vec{L}$. These possibilities will be further explored elsewhere. In the meanwhile, let us stress that the present potentials are local, except for their partial-wave dependence, which makes them quite different from usual realistic nucleon-nucleon interactions.

\section{Conclusions}

In conclusion, the method presented in this work can be considered as an optimal inversion technique for a given partial wave in the neutral case: it provides a minimal parameterization of the scattering phase shifts in terms of either scattering-matrix poles or
effective-range Padé expansion,
together with an analytical expression for the corresponding potential. The whole algorithm can readily be summarized by a short computer code \cite{code}. The only difficulty of the method is that the scattering-matrix poles sometimes become complex when the effective-range function is used as starting point for the inversion, which might lead to oscillating potentials.
A direct fit of the poles should then be performed,
with the double constraint of staying on the imaginary axis (except for possible resonances)
and of satisfying a well-defined effective-range expansion for $l>0$.
For the singlet neutron-proton case in the $S$ and $D$ waves,
the obtained poles and potentials are satisfactory.
The $S$-wave potential is shallow while the $D$-wave potential is deep,
which opens the way to a new discussion of the deep/shallow ambiguity in this case. The local nature of these potentials makes them quite different from other realistic nucleon-nucleon interactions but a comparison between different models could be made directly through their effective-range-function or scattering-matrix-poles properties. Further developments of the method might include the link between different partial waves, the comparison between the neutral and charged cases, generalization to the coupled-channel case \cite{pupasov:11,baye:14}, and application to elastic collisions in atomic physics \cite{blackley:14}.\\

\begin{acknowledgments}
This text presents research results
of the IAP program P7/12 initiated by the Belgian-state
Federal Services for Scientific, Technical, and Cultural Affairs.
The author B.M. is a beneficiary of a post-doctoral grant from the Belgian
Federal Science Policy Office (BELSPO) co-funded by the Marie-Curie Actions (FP7) from the European Commission.
\end{acknowledgments}


\begin{thebibliography}{19}
\bibitem{code} The Python code for the method is available on request to the authors.
\end{thebibliography}

\end{document}